\documentclass[12pt,preprint]{aastex}

\shorttitle{Interpretation of Comet 9P/Tempel 1}
\shortauthors{Yamamoto et al.}

\usepackage{graphicx}
\usepackage{fancybox}
\usepackage{color}
\fancyput(-1.in,-8.5in){
	\color[rgb]{0.85,0.85,0.85}{
		\rotatebox{50}{
			\scalebox{9}{ApJL 673, L199, 2008.}
		}
	}
}
\begin{document}


\title{COMET 9P/TEMPEL 1: INTERPRETATION WITH THE DEEP IMPACT RESULTS}


\author{Satoru Yamamoto}
\affil{Graduate school of frontier science, University of Tokyo, Kashiwanoha 5-1-5, Kashiwa, 277-8562, Japan}
\email{yamachan@gfd-dennou.org}
\author{Hiroshi Kimura, Evgenij Zubko, Hiroshi Kobayashi, Koji Wada}
\affil{Institute of Low Temperature Science, Hokkaido University, Kita-ku Kita--19 Nishi--8, Sapporo, 060-0819, Japan}
\email{hiroshi\_kimura@lowtem.hokudai.ac.jp, zubko@lowtem.hokudai.ac.jp, hkobayas@lowtem.hokudai.ac.jp, wada@lowtem.hokudai.ac.jp}
\author{Masateru Ishiguro}
\affil{Astronomy Program, Department of Physics \& Astronomy, Seoul National University, San 56-1, Sillim-dong, Gwanak-gu, Seoul 151-742, Korea}
\email{ishiguro@astro.snu.ac.kr}
\and
\author{Takafumi Matsui} \affil{Graduate school of frontier science, University of Tokyo, Kashiwanoha 5-1-5, Kashiwa, 277-8562, Japan} \email{matsui@k.u-tokyo.ac.jp.}





\begin{abstract}

According to our common understandings, the original surface of a short-period comet nucleus has been lost by sublimation processes during its close approaches to the Sun.
Sublimation results in the formation of a dust mantle on the retreated surface and in
chemical differentiation of ices over tens or hundreds of meters below the mantle.
In the course of NASA's Deep Impact mission, optical and infrared imaging observations of the ejecta plume were conducted by several researchers,
but their interpretations of the data came as a big surprise:
(1) The nucleus of comet 9P/Tempel 1 is free of a dust mantle, but maintains its pristine crust of submicron--sized carbonaceous grains;
(2) Primordial materials are accessible already at a depth of several tens of cm with abundant silicate grains of submicrometer sizes.
In this study, we demonstrate that a standard model of cometary nuclei explains well available observational data:
(1) A dust mantle with a thickness of $\sim 1$--$2$~m builds up on the surface, where compact aggregates larger than tens of micrometers dominate;
(2) Large fluffy aggregates are embedded in chemically differentiated layers as well as in the deepest part of the nucleus with primordial materials.
We conclude that the Deep Impact results do not need any peculiar view of a comet nucleus.

\end{abstract}



\section{INTRODUCTION}

Short-period comets are originally composed of ices and dust that were present in the cold comet-formation region of the primordial solar nebula.
When the comets leave the trans-Neptunian region and approach the Sun, volatile components of its nucleus begin to sublime.
Expanding volatile gases drag light dust having a large surface-to-volume ratio and leave preferentially heavy dust of a small surface-to-volume ratio on the surface of a comet to form a dust mantle \citep[see, e.g.,][for a review]{Prialnik}.
Indeed, optical and infrared observations of cometary comae suggest that large compact aggregates of tens of micrometers or larger lie in a dust mantle with depleted volatile materials, while large fluffy aggregates are embedded with volatile materials below the dust mantle \citep{Kolokolova}.
Dust mantle formation itself does not differentiate the composition of the aggregates, while chemical differentiation of a nucleus takes place owing to preferential sublimation losses of highly volatile materials in deeper parts of the nucleus.
Pristine amorphous water ices also crystallize to form a crystalline ice crust just below the dust mantle.
In addition, impact cratering by interplanetary bodies affects the surface evolution of the nucleus \citep[e.g.,][]{Fernandez}.
Therefore, primordial materials in comet nuclei remain only in the deepest region far below the dust mantle.
The above mentioned scenario for the evolution of a comet nucleus (hereafter referred to as the standard model)
is applicable to comet 9P/Tempel 1 (hereafter, T1), the target of NASA's Deep Impact (DI) mission:
Numerical studies on thermal evolution of the T1 nucleus predict the formation of a dust mantle with a thickness of $\sim 1$~m
and a crystalline ice crust with a thickness of $40$--$240$~m below the mantle \citep{Bar-Nun,Sanctis}.

Observations of T1 during the DI event gave us a good opportunity of examining the standard model for the T1 nucleus,
because DI excavated the T1 nucleus up to $\sim 10$--$20$ m depth \citep[e.g.,][]{Sunshine}.
Plenty of DI observational results support the standard model:
(1) The presence of a dust mantle on T1 is evidenced by a low thermal inertia of the nucleus estimated from its thermal maps \citep{A'Hearn};
(2) Near-infrared spectroscopic observations of volatiles excavated by the DI event revealed thermal processing of the layers below the dust mantle \citep{Mumma};
(3) The existence of large compact dust in the pre--impact T1 coma has been reported by many researchers \citep{Sykes,Farnham,Reach}.
The existence of such large dust in the pre--impact T1 is consistent with the standard model, because the dust mantle formation depletes only fluffy aggregates.

However, recent several researchers drew a peculiar picture of a comet nucleus from their observations of the ejecta plume produced by the DI event (hereafter, DI ejecta plume).
Subaru Telescope observation (STO) by \citet{STO-K} showed that the outer part of the DI ejecta plume (high-velocity ejecta) have a higher color temperature and lower infrared silicate-feature strengths than the inner part of the DI ejecta plume (low-velocity ejecta).
They attributed the high-velocity ejecta to submicron-sized carbonaceous grains and the low-velocity ejecta to submicron-sized silicate grains.
As a result, they proposed that the near-surface region of T1 is covered by a pristine crust of submicron-sized carbonaceous grains (hereafter, pristine carbon crust), which were created by cosmic rays in the trans-Neptunian region.
In addition, they interpreted the existence of many craters observed by \citet{Thomas} as evidence of a long surface retention age of the pristine carbon crust.
This means that the nucleus of a comet does not form a dust mantle during its close approaches to the Sun.
They also claim that the pristine carbon crust is very thin (several tens of cm) based on the ejection velocity and their estimate of the amount of carbonaceous grains.
Finally, they concluded that even at a depth of only several tens of cm below the surface comet materials keep their primordial composition together with submicron-sized silicate grains.
\citet{Furusho} supported this interpretation by measuring slightly higher linear polarizations for the high-velocity ejecta compared to the low-velocity ejecta.
\citet{Harker} also supported these interpretations from their infrared observations.
These conclusions (hereafter, KFH model) completely contradict our conventional images of comet nuclei based on the standard model.

Note that the difference in the color temperature between high- and low-velocity ejecta may simply reflect the amount of volatile materials, because color temperatures depend on the volatile content \citep{Mukai2}.
It is also worth noting that the silicate--feature strength is not a good indicator of an amount of silicate grains nor their sizes \citep{Okamoto};
Large fluffy aggregates of tens of micrometers or larger show strong silicate--features, while large compact aggregates of the same composition and mass show weaker silicate--features \citep{Kolokolova}.
Furthermore, the degree of linear polarization for aggregates may slightly vary with their porosity and/or volatile content \citep{Kimura06}.

In this study, we examine how well the standard model can explain the available observational data of the DI event by calculating color temperature, silicate--feature strength, and polarization for aggregates in the framework of the standard model.
We then discuss the advantage of the standard model over the KFH model even when comparing with the results of \citet{STO-S} who analyzed the same STO data as \citet{STO-K}
(hereafter, we refer \citet{STO-K} and \citet{STO-S} as to STO-K and STO-S, respectively).
We also discuss the groundless interpretations of the long retention age of the T1 primordial surface and the thickness of its dust mantle based on comprehensive observational and theoretical considerations.

\section{AGGREGATE MODEL}

We calculate the color temperature $T_{\rm c}$, the silicate--feature strength $S_{\rm si}$, and the degree of linear polarization $P_{\rm l}$ for the standard model as follows.
In the standard model, large compact aggregates exist in a dust mantle and large fluffy aggregates are embedded with volatiles below the dust mantle \citep{Kolokolova}.
To describe large aggregates, we use fractal clusters of $2^{22}$ identical monomers with a radius of $0.1~\micron$ \citep{Kolokolova}.
Each monomer is assumed to have a concentric structure of an organic refractory outer layer, a forsteritic inner layer, and an amorphous silicate core \citep{YamaT}.

The compact aggregates are assumed to have a fractal dimension $D=2.5$, which corresponds to the value when the maximum compression of fluffy aggregates is attained \citep{Wada}.
Our calculation using the superposition T-matrix method (TMM) showed that 
the geometric albedo of the compact aggregates is $\sim 0.04$ at visible wavelengths \citep[see][]{Kimura},
and this value is identical to the albedo derived from an observation of the T1 surface (A'Hearn et al. 2005).
We also calculate $\beta \sim 0.3$ (the ratio of solar radiation pressure to solar gravity) for the compact aggregates, 
using Mie theory with the optical constants deduced from the Maxwell-Garnett mixing rule (MG) \citep[see][]{Mukai}:
This is also consistent with $\beta$ for high-velocity ejecta within the DI ejecta plume \citep[e.g.,][]{Meech05}.

Similarly, let large fluffy aggregates embedded with volatile materials below the dust mantle be presented by fractal clusters with $D=1.9$ \citep{Mukai}.
\citet{Sunshine} found the $3\,\micron$ feature of solid H$_2$O ice in their observations of the DI ejecta plume.
Sublimation of volatiles within the ejecta seems to have lasted over the course of a few days \citep{Schleicher}.
We thus consider ongoing sublimation of volatiles in the fluffy aggregates and assume that a few percent of volatiles in volume still remain on the surface of the aggregates.
(For a comparison, we also consider volatile--free fluffy aggregates).

We then derive the absorption cross sections $C_{\lambda}$ for the compact and fluffy aggregates, using Mie theory with the optical constants deduced from the MG \citep[e.g.,][]{Mukai}.
The use of the MG for aggregates has been examined based on the discrete dipole approximation (DDA) \citep{Kolokolova}.
From values of $C_{\lambda}$ at wavelengths $\lambda$ of $8.8$, $10.5$, and $12.4\,\micron$ where STO-K observed, we calculate the color temperature as $T_{\rm c}=C_{8.8}/C_{12.4}$
and the silicate-feature strength as $S_{\rm si}=2C_{10.5}/(C_{8.8}+C_{12.4})-1$.
We also calculate $P_{\rm l}$ at $\lambda = 0.8\,\micron$ and at a phase angle of the DI event for these aggregates.
Since $P_{\rm l}$ does not strongly depend on the number of monomers, we perform numerical calculations with clusters of 256 monomers by using the DDA and the TMM.
All the results are listed in Table 1.

\section{STANDARD INTERPRETATION}

We first interpret the observational data by STO-K.
Table 1 shows that the compact aggregate has higher $T_{\rm c}$ and lower $S_{\rm si}$ than the volatile--coated fluffy aggregate.
Impact cratering mechanism indicates that high-velocity ejecta are thrown from near--surface regions 
and low-velocity ejecta contain primarily materials excavated from deeper regions \citep[e.g.,][]{Melosh}.
Thus, a natural expectation in the standard model is that high-velocity ejecta in the DI ejecta plume would have higher $T_{\rm c}$ and lower $S_{\rm si}$ than the low-velocity ejecta.
This agrees with the observational data by STO-K that show the high-velocity ($\sim 180$ m s$^{-1}$) ejecta to have higher $T_{\rm c}$ and lower $S_{\rm si}$ than the low-velocity ejecta ($\sim 100$ m s$^{-1}$).

We next compare the polarimetric observation by \citet{Furusho} with $P_{\rm l}$ in Table 1. 
Table 1 shows that the both aggregates have a positive polarization  and that $P_{\rm l}$ for the compact aggregates (the high--velocity ejecta) is larger than that for the fluffy aggregates (the low--velocity ejecta).
This is consistent with the observation by \citet{Furusho} who showed that the high--velocity ejecta have a slightly higher $P_{\rm l}$ compared to the low--velocity ejecta.

\begin{figure}[htbp]
\begin{center}
\plotone{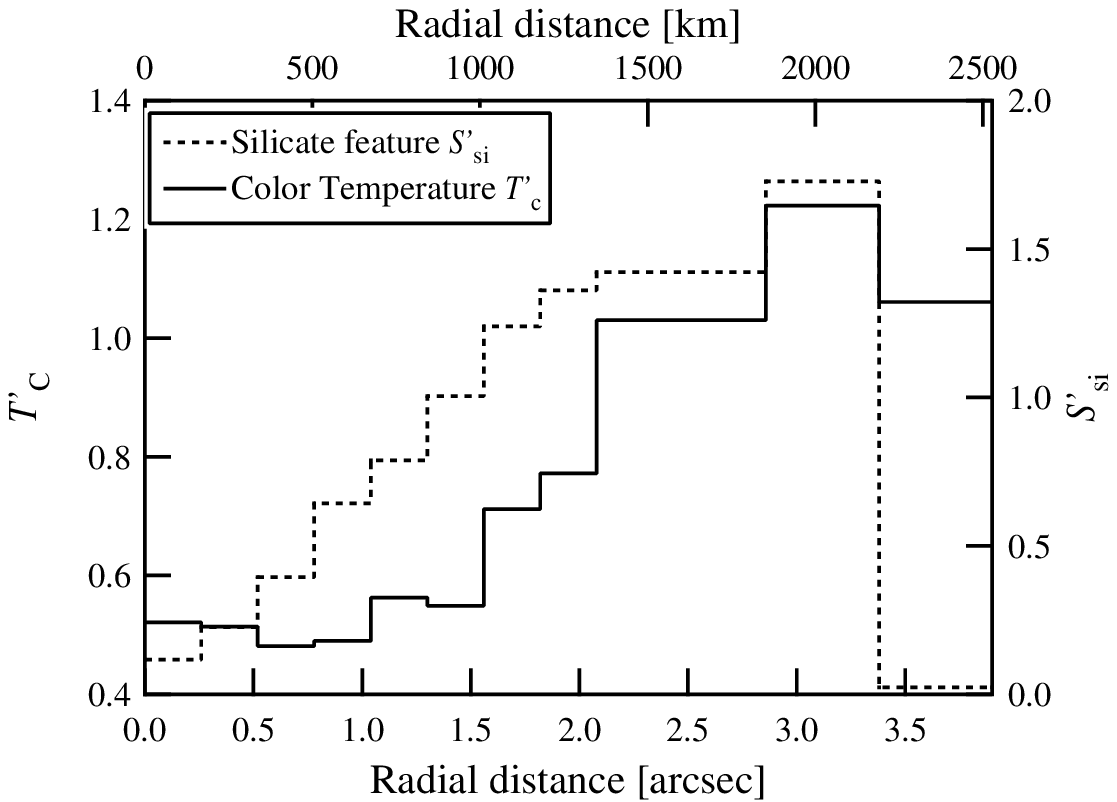}
\caption{The radial distributions of the color temperature $T^{\prime}_{\rm c}$ and the silicate--feature $S^{\prime}_{\rm si}$ derived from the data of Sugita et al. (2005; STO-S) at $t=3.5$ hrs.
The radial distance is the value from the nucleus whose actual position is located somewhere in the point spreading function ($\sim 0$\farcs$4$) of STO (see, STO-S).}
\label{f:1}
\end{center}
\end{figure}

The STO data at time $t=3.5$~hrs after the DI event are shown by STO-S as infrared spectra at various radial distances.
We also interpret the STO-S data by estimating the radial distributions of $T^{\prime}_{\rm c}$ and $S^{\prime}_{\rm si}$ from the data, where $T^{\prime}_{\rm c} = I_{8.8}/I_{12.4}$, $S^{\prime}_{\rm si}=2I_{10.5}/(I_{8.8}+I_{12.4})-1$, $I_{\lambda}$ being the STO intensity at a wavelength $\lambda$ of $8.8$, $10.5$, or $12.4\,\micron$, respectively.
Figure~\ref{f:1} shows that both $T^{\prime}_{\rm c}$ and $S^{\prime}_{\rm si}$ at $t=3.5$~hrs have the same peak position at $\sim 3$\farcs$0$.
We consider that the peaks represent low--velocity, volatile--free fluffy aggregates, because
(1) the peak position is within the expected position ($\ga 2$\farcs$5$) of the low-velocity ejecta at $t=3.5$~hrs;
(2) the peak position of the high-velocity ejecta at $t=2.0$~hrs is already $2$\farcs$5$ (STO-K), so that the expected position at $t=3.5$~hrs is beyond $\sim 4$\farcs$0$;
(3) the volatiles in the fluffy aggregates would sublime gradually through a few hours, resulting in higher $T_{\rm c}$ (see Table 1).
Thus, the STO data at $t=3.5$~hrs is well explained by the standard model.

In summary, the optical properties of the compact and fluffy aggregates based on the standard model well account for the observational data used by STO-K, \citet{Furusho}, and STO-S.

\section{DISCUSSION}

In this section, we refute the following interpretations in the KFH model: 
(1) the presumption of the pristine carbon crust;
(2) the long retention age of the pristine carbon crust;
(3) the existence of cometary primordial materials at a depth of several tens of cm from the surface.

\subsection{Peculiar Pristine Carbon Crust}

In the KFH model, submicron--sized ($0.3\,\micron$ radius) carbonaceous grains are responsible to the $T^{\prime}_{\rm c}$ peak within the high--velocity ejecta ($\sim 180$ m s$^{-1}$).
If this were correct, the carbonaceous grains in the high-velocity ejecta must have been decelerated significantly through $t=2$--$3.5$~hrs, because the peak position of $T^{\prime}_{\rm c}$ is $\sim 2$\farcs$5$ at $t=2$~hrs (STO-K) and proceeds to only $\sim 3$\farcs$0$ at $t=3.5$~hrs (Fig.~\ref{f:1}), corresponding to an average velocity of $60$~m s$^{-1}$.
Furthermore, STO-K claimed that the high-velocity ejecta ($\ga 160$~m s$^{-1}$) were observed at $t=6$--$9$ hrs by \citet{Furusho} as highly polarized regions at $\ga 6^{''}$.
Therefore, the KFH model suggests that the high-velocity ($\sim 180$ m s$^{-1}$) carbonaceous grains were suddenly decelerated to $\sim 60$~m s$^{-1}$ during $t=2$--$3.5$ hrs and then accelerated again to $\ga 160$~m s$^{-1}$ after $t=3.5$~hrs.
We believe that no one could support such a dynamical behavior.
A most likely explanation of the STO data is that the high-velocity ejecta expand with a nearly constant velocity over the first 20 hrs as observed by \citet{Meech05} 
and that low-velocity ejecta is responsible to the $T^{\prime}_{\rm c}$ peak at $t=3.5$~hrs.
In addition, the observational constraint of $\beta \sim 0.3$ in the high-velocity ejecta completely rules out the predominance of $0.3\,\micron$ carbonaceous grains, which have $\beta > 1$ \citep{Burns}.
Therefore, the pristine carbon crust is a false interpretation, unless the above contradictions are resolved.

\subsection{Crater Formation on the T1 Surface}

STO-K interpret the existence of craters with radii up to $\sim 200$\,m \citep{Thomas} as evidence for a long surface retention age of the pristine carbon crust, because they assumed that observable craters are produced when T1 was in the trans-Neptunian region.
However, \citet{Kadono} himself estimated that a comet collides Near-Earth Asteroids (NEAs) of 1\,m every $\sim100$ years during its approaches the Sun,
forming a $\sim100$\,m sized crater.
Short-period comets are also bombarded in the main asteroid belt with higher collisional probabilities than those of NEAs \citep[e.g.,][]{Hawkes}.
The model calculation by \citet{Gronkowski} shows that the number of observable craters on T1 is around 10 for the steady state of collisional environment in the main asteroid belt and up to $\sim 120$.
Even if we use a different parameter set with the asteroidal collision model by \citet{Gil--Hutton} and \citet{Hawkes}, the number of $\sim 200$\,m-sized craters produced at each orbit is $\sim 20$.
Thus, many craters with radii $\la 200$\,m can be formed during every approaches to the Sun.
Furthermore, sublimation processes increase crater sizes with time \citep{Britt},
suggesting that the craters observed on T1 may have been produced by smaller impactors with more frequent impact rates.
We thus conclude that the existence of craters on T1 cannot be attributed to a long surface retention age.
 
\subsection{Thickness of the T1 Dust Mantle}

The standard model predicts the thickness of the dust mantle on T1 to be several meters or thinner \citep{Sanctis}.
Following STO-K, we could estimate the maximum depth of the region from which the compact aggregates are excavated, by using the $Z$-model \citep{Maxwell}.
This model allows us to estimate the maximum depth of an excavation flow (maximum excavation depth) for each ejecta with ejection velocity $v_{\rm e}$.
As a result, we might arrive at the dust mantle thickness of several tens of cm as STO-K, but we find several problems in their assumptions.
According to \citet{Croft} and \citet{Melosh}, the center of an excavation flow (flow center) in the $Z$--model is at some depth below the surface,
which may be comparable to a projectile diameter even for oblique impacts.
However, STO-K assume that the flow center is at the surface.
This is not correct: the actual excavation depth is the sum of the depth of flow center (the projectile diameter, $\sim 1$~m) and  the maximum excavation depth.
In addition, an acceleration of the ejecta due to the gas plume observed by STO-S was not taken into account.
Then, the value of $v_{\rm e}$ used by STO-K is overestimated and the resulting thickness is underestimated.
Although STO-K claim that the thickness of several tens of cm is similar to that derived from the total mass of carbonaceous grains in the DI ejecta plume, 
the presumption of the pristine carbon crust is a false interpretation, invalidating their estimate of the total mass of the carbonaceous grains.
As a conservative estimate, the DI results suggest that the thickness of the dust mantle is $1$--$2$\,m.

\acknowledgments

This study was supported in part by T. Yoda, a Grant-in-Aid for Scientific Research from Japan Society for the promotion of Science, 
and a Grant-in-Aid for Scientific Research on Priority Areas from Ministry of Education, Culture, Sports, Science and Technology, Japan.

\clearpage

\clearpage

\begin{table}
\begin{center}
\caption{Color temperature $T_{\rm c}$, silicate--feature strength $S_{\rm si}$, and linear polarization $P_{\rm l}$.}
\begin{tabular}{cccc}
\tableline\tableline
& $T_{\rm c}$ & $S_{\rm si}$ & $P_{\rm l}$ \\
\tableline
Compact aggregate   & $1.94$ & $3.08$ & $0.211$ \\
Volatile-coated fluffy aggregate & $1.62$ & $3.17$ &$0.167$ \\
Volatile-free fluffy aggregate      & $1.99$ & $3.51$ & $0.187$ \\

\tableline
\end{tabular}
\end{center}
\end{table}

\end{document}